\documentclass[aps,prd,preprint,superscriptaddress,showpacs,floatfix,
preprintnumbers,nofootinbib, a4paper]{revtex4-1}
\usepackage{color}
\usepackage{graphicx}
\usepackage{textcomp}
\usepackage{amsmath,amssymb}
\usepackage{enumerate}
\usepackage{multirow}
\newcommand{\mathsym}[1]{{}}

\newcommand{\ba}{\begin{array}} 
\newcommand{\ea}{\end{array}}
\newcommand{\be}{\begin{equation}}
\newcommand{\ee}{\end{equation}}
\newcommand{\beqa}{\begin{eqnarray}} 
\newcommand{\eeqa}{\end{eqnarray}}
\def\321{$SU(3)\times SU(2)\times U(1)$}
\def\mt{$\mu$-$\tau$~} 
\def\vev#1{\left\langle #1\right\rangle}
 
\newcommand{\Dms}{\Delta m^2_{\rm sol}}
\newcommand{\Dma}{\Delta m^2_{\rm atm}}

\begin{document} 
\vspace*{1cm}
\title{Generalized $\mu$-$\tau$ symmetry and discrete subgroups of $O(3)$} 
\bigskip 
\author{Anjan S. Joshipura}
\email{anjan@prl.res.in}
\affiliation{Physical Research Laboratory, Navarangpura, Ahmedabad 380 009, India.} 
\author{Ketan M. Patel}
\email{ketan.patel@pd.infn.it} 
\affiliation{Istituto Nazionale Fisica Nucleare, Sezione di Padova, I-35131 Padova, Italy. \vspace*{1cm}}

\begin{abstract}
The generalized $\mu$-$\tau$ interchange symmetry in the leptonic mixing matrix
$U$ corresponds to the relations: $|U_{\mu i}|=|U_{\tau i}|$ with $i=1,2,3$.
It predicts maximal atmospheric mixing and maximal Dirac CP violation given
$\theta_{13} \neq 0$. We show that the generalized $\mu$-$\tau$ symmetry can
arise if the charged lepton and neutrino mass matrices are invariant under
specific residual symmetries contained in the finite discrete subgroups of
$O(3)$. The groups $A_4$, $S_4$ and $A_5$ are the only such groups which can
entirely fix $U$ at the leading order. The neutrinos can be (a) non-degenerate
or (b) partially degenerate depending on the choice of their residual
symmetries. One obtains either vanishing or very large $\theta_{13}$ in case of
(a) while only $A_5$ can provide $\theta_{13}$ close to its experimental value
in the case (b). We provide an explicit model based on $A_5$ and discuss a class
of perturbations which can generate fully realistic neutrino masses and mixing
maintaining the generalized $\mu$-$\tau$ symmetry in $U$. Our approach provides
generalization of some of the ideas proposed earlier in order to obtain the
predictions, $\theta_{23}=\pi/4$ and $\delta_{\rm CP} = \pm \pi/2$.
\end{abstract} 

\maketitle

\section{Introduction} 
\label{intro}
The data from various neutrino oscillation
experiments analyzed in the context of three neutrino oscillations have revealed
five fundamental parameters by now
\cite{Gonzalez-Garcia:2014bfa,Capozzi:2013csa,Forero:2014bxa}. These include two
squared differences of neutrino masses and three mixing angles in the
Pontecorvo-Maki-Nakagawa-Sakata mixing matrix $U_{\rm PMNS}$. For any of the
normal or inverted ordering in the neutrino masses, their $3\sigma$ ranges can
be summarized as \cite{Gonzalez-Garcia:2014bfa}:
$$
0.270<\sin^2\theta_{12}<0.344~,~~~0.385<\sin^2\theta_{23}<0.644~,~~~0.0188<\sin^2\theta_{13}<0.0251$$
$$7.02 < \frac{\Delta m^2_{21}}{10^{-5}~{\rm eV}^2} < 8.09~,~~2.325 <
\frac{\Delta m^2_{31}}{10^{-3}~{\rm eV}^2} < 2.599~{\rm or}~-2.259 <
\frac{\Delta m^2_{32}}{10^{-3}~{\rm eV}^2} < -2.307$$ 
Here $\Delta m^2_{ij}
\equiv m_i^2-m_j^2$. Discrete symmetry based approaches have been quite widely
used in order to explain the special values of lepton mixing angles, see for
example recent reviews
\cite{Altarelli:2010gt,Altarelli:2012ss,Smirnov:2011jv,King:2013eh,
Ishimori:2010au}.
One assumes that the global symmetry group $G_f$ of the leptons is spontaneously
broken to the smaller symmetries $G_\nu$ and $G_l$ of the neutrino and the
charged lepton mass matrices respectively. The leptonic mixing can solely be
fixed from the choice of $G_l$ and $G_\nu$ in a given $G_f$
\cite{Lam:2007qc,Lam:2008rs,Lam:2008sh,Lam:2012ga,Lam:2011ag}. Possible choices
of $G_f$ leading to three non-degenerate neutrinos are extensively studied in
\cite{Toorop:2011jn,deAdelhartToorop:2011re,Hernandez:2012ra,Hernandez:2012sk,
Holthausen:2012wt,Parattu:2010cy,Fonseca:2014koa,Talbert:2014bda} and mixing
patterns are
analyzed. In a novel approach, it is shown that suitable choices of $G_f$ can
also lead to the cases with one massless neutrino
\cite{Joshipura:2013pga,Joshipura:2014pqa}, two or three degenerate neutrinos
\cite{Hernandez:2013vya,Joshipura:2014qaa} and two degenerate and one massless
neutrino \cite{Joshipura:2014qaa}. In an alternate approach, it is shown
recently in \cite{Joshipura:2015zla} that a massless neutrino with/without a
degenerate pair of neutrinos can arise if neutrino mass matrix is assumed to be
anti-symmetric under $G_\nu$.

After the clear evidence of nonzero $\theta_{13}$ and with the most recent data,
we now start to have an indirect indication of the sixth parameter, namely the Dirac
CP phase $\delta_{\rm CP}$ in the lepton sector. In fact the observed value of
$\theta_{13}$ and measured combination of $\theta_{13}$ and $\delta_{\rm CP}$ by
T2K long-baseline experiment \cite{Abe:2013hdq} are in good agreement if
$\delta_{\rm CP} \sim -\pi/2$ \cite{Strumia:2015dca,Elevant:2015ska}. This however is a mere
indication at present and more data will certainly provide clear picture in the
near future. Nevertheless, such a special value of CP phase may be indicative
signal of some hidden symmetries in the lepton sector. The current global fits of
neutrino oscillation data disfavours the maximal atmospheric mixing angle at
$1\sigma$ however it is in accordance with the data at $3\sigma$ in case of both
normal and inverted ordering in the neutrino masses. The ansatz and symmetries
of neutrino mass matrix predicting $\theta_{23}=\pi/4$ and $\delta_{\rm CP} =
\pm \pi/2$ have been proposed earlier in
\cite{Ma:2001dn,Harrison:2002et,Babu:2002dz,Grimus:2003yn}. In the simplest
case, the above prediction can be obtained if the Majorana neutrino mass matrix
in the diagonal basis of the charged leptons, namely ${\cal M}_{\nu f}$,
satisfies \be \label{gmt} S_{23}^T {\cal M}_{\nu f} S_{23} = {\cal M}_{\nu f}^*,
\ee where \be\label{s23} S_{23}=\left( \ba{ccc} 1 & 0& 0\\ 0 & 0 & 1\\ 0 & 1 & 0
\ea \right)~.\ee The symmetry transformation is a discrete $Z_2$ symmetry
corresponding to \mt interchange together with CP conjugation
\cite{Grimus:2003yn,Kitabayashi:2005fc,Farzan:2006vj,Joshipura:2009tg,
Gupta:2011ct,Grimus:2012hu,Mohapatra:2012tb,Feruglio:2012cw,Holthausen:2012dk,Nishi:2013jqa,Mohapatra:2015gwa}.
Such an ${\cal M}_{\nu f}$ leads to the relations among the elements of PMNS
matrix, $U \equiv U_{\rm PMNS}$: 
\be \label{gen-pred} |U_{\mu i}|=|U_{\tau
i}|~
~{\rm for}~ i=1,2,3~\ee 
and predicts $\theta_{23} = \pi/4$ and
$\sin\theta_{13} \cos\delta_{\rm CP}=0$, equivalently $\delta_{\rm CP} = \pm
\pi/2$ if $\theta_{13} \neq 0$. The relations in Eq. (\ref{gen-pred}) were first
proposed in \cite{Harrison:2002et} and we refer them as the results of
``generalized \mt symmetry" in the leptonic mixing matrix\footnote{This is also
referred as ``\mt reflection symmetry" in some literature
\cite{Harrison:2002et,Farzan:2006vj}.}.

We show that predictions in Eq. (\ref{gen-pred}) or generalized \mt symmetry
arise on more general grounds and can follow without invoking CP and/or the \mt
symmetry and can follow even if Eq. (\ref{gmt}) is not satisfied by ${\cal
M}_{\nu f}$. As we shall see, Eq. (\ref{gen-pred}) arises if neutrinos and the
charged lepton mass matrices are invariant under specific residual symmetries
contained in some discrete subgroups (DSG) of $O(3)$. The residual symmetries
contained in DSG of $O(3)$ can be used to get a neutrino mass matrix with
non-degenerate or partially degenerate spectrum with two of the masses being
equal. The generalized \mt symmetry follows in both the cases. While the general
result that we derive holds for any DSG of $O(3)$, we shall discuss specific
examples of groups having three dimensional irreducible representation (irreps).
There are only three such groups, namely $A_4$, $S_4$ and $A_5$. All of which have
been widely discussed in the literature
\cite{Altarelli:2010gt,Altarelli:2012ss,Smirnov:2011jv,King:2013eh,Ishimori:2010au}
and we shall recapitulate some of the known results and present new examples
specifically in case of the partially degenerate neutrino mass spectrum.

The $A_5$ symmetry together with CP transformation has been studied recently in
\cite{Everett:2015oka,Li:2015jxa,DiIura:2015kfa,Ballett:2015wia} in order to
predict the
neutrino mixing angles and CP phases in the case of three massive Majorana
neutrinos. Our approach is different from these works as we do not impose CP explicitly
but discuss situations under which the generalized CP predictions arise
automatically. Also the choice of residual symmetry $G_\nu$ leading to
degenerate solar pair is not considered in the quoted works.

In the next section, we present our main result and discuss the emergence of
generalized \mt symmetry from the DSG of $O(3)$. We then discuss specific
examples of the general result in Section \ref{example}. An explicit model based
on the $A_5$ group is constructed in the Section \ref{model}. Finally, we
summarize in the last section.

\section{Discrete subgroups of $O(3)$ and maximal $\theta_{23}$ \& $\delta_{\rm
CP}$} 
\label{review}
We discuss the sufficiency conditions leading to
generalized \mt symmetry predictions, Eq. (\ref{gen-pred}). Let $T_l$, $T_\nu$
and $S_{a\nu}$ with $a=1,2$ denote $3\times 3$ {\it real} orthogonal matrices
with the property \be\label{con} T_l^n=T_\nu^m=S_{a\nu}^2=1~ ~{\rm with}~
n,m\geq 3~\ee and $[S_{1\nu},S_{2\nu}]=0$, $[T_l,T_\nu]\not=0$,
$[T_l,S_{a\nu}]\not=0$. Let the Hermitian combination $M_lM_l^\dagger$ of the
charged lepton mass matrix $M_l$ satisfy 
\be \label{cl-invariance} T_l^\dagger
M_l M_l^\dagger T_l = M_l M_l^\dagger~\ee 
and neutrino mass matrix be invariant
under either $S_{a\nu}$ or $T_\nu$:
\be \label{cn-invariance} {\rm (a)} ~~
S_{a\nu}^T M_\nu S_{a\nu}=M_\nu ~~~~{\rm or}~~~~{\rm (b)}~~ T_\nu^TM_\nu
T_\nu=M_\nu~.\ee Then the resulting $U_{\rm PMNS}$ displays the exact
generalized \mt symmetry with elements satisfying Eq. (\ref{gen-pred}). It is
clear that if $(T_l,~S_{a\nu})$ or $(T_l,~T_\nu)$ close to a finite group, then
the minimal such group would be a DSG of $O(3)$. Thus the DSG of $O(3)$ can
naturally lead to the generalized \mt symmetry.

The case (a) in Eq. (\ref{cn-invariance}) corresponds to three
non-degenerate
neutrino masses and (b) to partially degenerate spectrum with two equal
neutrino masses. The neutrino mass matrix is invariant under a $Z_2\times Z_2$
symmetry in the case (a). This symmetry corresponds to changing the signs of any
two of the three neutrino fields in their mass basis. Such a symmetry is always
present if all three neutrinos are massive Majorana particles and
non-degenerate. If two
of the neutrinos are degenerate then the residual symmetry is bigger since one
can multiply the corresponding fields $\nu_1$ and $\nu_2$ by complex phase $\eta$
and $\eta^*$ respectively leaving their combined mass term invariant. The
residual symmetry in this case is $Z_m$ with $m\geq 3$ and implies a partially
degenerate spectrum which has been considered in detail in
\cite{Hernandez:2013vya,Joshipura:2014qaa}.

The proof of the above uses an important and well known result that matrices
diagonalizing symmetry operators of the mass matrices also diagonalize the
corresponding mass matrices themselves
\cite{Lam:2007qc,Lam:2008rs,Lam:2008sh,Lam:2012ga,Lam:2011ag}. Specifically, let
$V_l$ $(V_\nu)$ be $3\times 3$ unitary matrix diagonalizing the symmetry
operators $T_l$ ($S_{a\nu}$ or $T_\nu$). Then the matrices $U_l$ and $U_\nu$, diagonalizing $M_l M_l^\dagger$ and $M_\nu$ respectively, are
given by $U_l=V_l P_l$ and $U_\nu=V_\nu P_\nu$, where $P_l$ and $P_\nu$ are
arbitrary diagonal phase matrices. As a result, the elements of the 
$U \equiv U_{\rm PMNS}$ matrix satisfy
\be \label{UMNS}| U_{ij}| = | (U_l^\dagger U_\nu)_{ij}|=| (V_l^\dagger
V_\nu)_{ij}|~. \ee Eqs. (\ref{cl-invariance},\ref{cn-invariance}) allow us to
determine the general form of $V_l$ and $V_\nu$. For this, we note that
eigenvalues of any unitary matrix satisfies 
\be \label{ev-su3} \lambda^3-\chi
\lambda^2+\chi^*\lambda -1=0~, \ee
where $\chi$ denotes the trace of the matrix (or character)
and all the eigenvalues $\lambda$ satisfy $|\lambda|=1$. If $\chi$ is real then
one of the roots of the above equation is $\lambda_1=1$ and the other two are
given by $\lambda_{2,3}=\frac{1}{2}\left( \chi-1\pm \sqrt{(\chi-1)^2-4}\right)$.
This has only two real solutions of modulus one corresponding to $\chi=3$ and
$\chi=-1$. These respectively correspond to an identity element and elements of
order 2. The remaining solutions are non-real and complex conjugate to each
other. Such elements necessarily have order $\geq 3$. It follows that the
matrices $T_l$, $T_\nu$ satisfying Eq. (\ref{con}) have eigenvalues $\lambda_i =
(1, \eta, \eta^*)$ with $\eta \neq \pm 1$ and $|\eta|^2=1$ while $S_{a\nu}$ have
eigenvalues $(1,-1,-1)$.

Any $T_l$ with a pair of complex conjugate eigenvalues is necessarily
non-diagonal in the basis in which it is real and its eigenvalue equation is
given by 
\be \label{Tl-ev} T_l v_i=\lambda_i v_i~, \ee 
where $v_i$ are
eigenvectors. It follows from the eigenvalues of $T_l$ that $v_1$ can be chosen
real and $v_2 = v_3^*$. Thus, $V_l$ diagonalizing $T_l$ can be chosen to have a
general form 
\be\label{gen-vl} V_l=\left( \ba{ccc} x_1&z_1&z_1^*\\
x_2&z_2&z_2^*\\ x_3&z_3&z_3^*\\ \ea \right)~,\ee 
with real $x_i$ and complex
$z_i$. The corresponding matrix diagonalizing $M_lM_l^\dagger$ would be given by
$U_l=V_l P_l$. Next, we show that the matrix $U_\nu$ diagonalizing $M_\nu$ has
the form \be \label{gen-vnu} U_\nu = O_\nu Q_\nu~\ee in both the cases (a) and
(b), where $O_\nu$ is a real orthogonal matrix and $Q_\nu$ is a diagonal phase
matrix. Since $[S_{1\nu},S_{2\nu}]=0$, both $S_{a\nu}$ are diagonalized by a
common unitary matrix and since $S_{a\nu}$ and their eigenvalues are real, the
eigenvectors of $S_{a\nu}$ can also be chosen real. The same $O_\nu$ would
diagonalize the neutrino mass matrix also due to symmetry relation Eq.
(\ref{cn-invariance}). But the neutrino masses can be complex and $Q_\nu$ in Eq.
(\ref{gen-vnu}) corresponds to their phases. For the case (b), the matrix
$V_\nu$ that diagonalizes $T_\nu$ is formally the same as Eq. (\ref{gen-vl})
which diagonalizes $T_l$. This follows from the fact that both $T_l$ and $T_\nu$
are real and have a pair of complex conjugate eigenvalues. Thus we can write %
\be \label{vnu} V_\nu=\left( \ba{ccc} u_1&u_1^*&w_1\\ u_2&u_2^*&w_2\\
u_3&u_3^*&w_3\\ \ea \right)\ee with $w_i$ real. Note that the ordering of
eigenvectors is not determined from the symmetry arguments and we have chosen an
ordering in Eq. (\ref{gen-vl}) which would give generalized \mt symmetry. Other
choices would correspond to $e$-$\tau$ or $e$-$\mu$ symmetries leading to the
predictions $|U_{ei}|=|U_{\mu i}|$ or $|U_{ei}|=|U_{\tau i}|$ respectively in
$U_{\rm PMNS}$. The ordering in $V_\nu$ in Eq. (\ref{vnu}) is however chosen
requiring that the degenerate pair of neutrinos corresponds to the solar
neutrinos pair. While $V_\nu$ diagonalizing $T_\nu$ is given above, the
diagonalizing matrix $U_\nu$ does not differ from it merely by a phase matrix as
in the case of non-degenerate neutrinos. The degeneracy in the first two masses
implies \cite{Joshipura:2014qaa}
\be \label{unu} U_\nu = V_\nu
U_{12}R_{12}(\theta_X)P_{\beta_2}~, \ee with \be \label{u12} U_{12}=\left(
\ba{ccc} \frac{i}{\sqrt{2}}&\frac{1}{\sqrt{2}}&0\\
\frac{-i}{\sqrt{2}}&\frac{1}{\sqrt{2}}&0\\ 0&0&1\\ \ea \right)~,\ee
$R_{12}$
denoting arbitrary rotation in the 1-2 plane by an angle $\theta_{X}$ and
$P_{\beta_2}={\rm Diag.}~(1,1,e^{i\beta_2/2})$. It then follows from Eqs.
(\ref{unu},\ref{u12}) that $U_\nu$ also has the same form as given by Eq.
(\ref{gen-vnu}). It is then straightforward to verify that $U_l=V_l P_l$ with
$V_l$ as in Eq. (\ref{gen-vl}) and $U_\nu$ as in Eq. (\ref{gen-vnu}) lead to
$U_{\rm PMNS}$ matrix satisfying Eq. (\ref{gen-pred}).

A neutrino mass matrix which is $Z_2\times Z_2$ symmetric can in general possess
non-trivial phases represented by $Q_\nu$ in Eq. (\ref{gen-vnu}). If these
phases are trivial and if $U_l$ is in the form of Eq. (\ref{gen-vl}) then the
Majorana neutrino mass matrix in the diagonal basis of the charged leptons is
given by 
\be \label{mnf} {\cal M}_{\nu f}\equiv U_l^T M_\nu U_l=\left(
\ba{ccc} X & A & A^*\\ A & B & C\\ A^*&C&B^*\\ \ea \right)~,\ee 
where $X$ and
$C$ are real parameters. This provides the most general solution of Eq.
(\ref{gmt}). The above $M_{\nu f}$ was first obtained
\cite{Ma:2001dn,Babu:2002dz} in the context of $A_4$ model with quasidegenerate
neutrinos. It was then argued in \cite{Grimus:2003yn} that this form can result
from a combined operation of the \mt and CP symmetry and leads to prediction of
the maximal $\delta_{\rm CP}$.

If $M_\nu$ is $Z_2\times Z_2$ symmetric but Majorana phases are non-trivial then
even with $U_l$ as in Eq. (\ref{gen-vl}) one does not get the above specific
form of Eq. (\ref{mnf}) but Eq. (\ref{gen-pred}) still holds. Thus
the combined
operation of CP and \mt symmetry is sufficient but not necessary to get the the
maximal $\theta_{23}$ and $\delta_{\rm CP}$.

It has been noticed before \cite{Ferreira:2012ri,Ma:2015gka,Ma:2015pma} that the
from given in Eq. (\ref{mnf}) follows if $V_l$ is given by \be \label{uw}
V_l=U_\omega=\frac{1}{\sqrt{3}}\left( \ba{ccc} 1&1&1\\ 1&\omega&\omega^2\\
1&\omega^2&\omega\\ \ea \right)~\ee with $\omega=e^{2\pi i/3}$ and if neutrino
mass matrix is real. The above form of $V_l$ is a special case of our general
form, Eq.(\ref{gen-vl}) and results when $T_l$ is identified with a $Z_3$ group
associated with cyclic permutations of three objects. A similar case is also
studied recently in the contexts of type II seesaw
\cite{He:2015afa,He:2015gba,Dev:2015dha}.

We end this section with some important remarks connected with the above result.
\begin{itemize} 
\item If one were to replace $Z_n$ invariance of
$M_lM_l^\dagger$ also by a $Z_2\times Z_2$ symmetry then both $U_l$ and $U_\nu$
would be real upto a diagonal phase multiplication on right and $\delta_{\rm
CP}$ would be zero. If $Z_2\times Z_2$ invariance of $M_\nu$ in case of the 
non-degenerate neutrinos is replaced by a single $Z_2$ then reality of $V_\nu$ and
hence the prediction of the generalized \mt symmetry does not hold. An example
of this is found in a specific model \cite{Varzielas:2013hga} based on the $A_5$
group which uses a single $Z_2$ symmetry for neutrinos. As far as the degenerate
neutrinos are concerned, the order of $T_\nu$ is necessarily $>2$. Thus
all DSG of $O(3)$ giving degenerate neutrinos necessarily also give Eq.
(\ref{gen-pred}). 
\item If neutrinos are degenerate then both the solar
angle
and $\delta_{\rm CP}$ are undefined. This is reflected by the presence of the
unknown angle $\theta_X$ in Eq. (\ref{unu}). But note that the relations in Eq.
(\ref{gen-pred}) hold even if $U \to U R_{12}(\theta_X) P_{\beta_2}$ and
therefore the
arbitrariness in defining $U_\nu$ arising from
the degeneracy of the solar pair does not affect the undelying generalized
\mt symmetry. Equivalently, one finds
\cite{Hernandez:2013vya,Joshipura:2014qaa} that the quantity $I_\alpha \equiv
{\rm Im}(U_{\alpha 1}^* U_{\alpha 2})$ remains invariant under $U \to U
R_{12}(\theta_X) P_{\beta_2}$. These quantities can be written in the standard
parameterization of $U_{\rm PMNS}$ as 
\beqa \label{invariant}
c_{12}s_{12}\sin\frac{\beta_1}{2}&=&\frac{1}{c_{13}^2} I_e~,\nonumber \\
c_{12}^2\sin \left(\delta_{\rm CP}-\frac{\beta_1}{2} \right)+s_{12}^2\sin
\left(\delta_{\rm CP}+\frac{\beta_1}{2} \right)&=&
\frac{1}{s_{23}c_{23}s_{13}}\left(I_\mu-\frac{s_{23}^2s_{13}^2-c_{23}^2}{c_{
13}^2}I_e\right),\eeqa
where $s_{ij}=\sin\theta_{ij}$ and
$c_{ij}=\cos\theta_{ij}$. Using the form of $U_{\rm PMNS}$ obtained in the
degenerate case above, one finds that $I_e =0$ and $I_\mu = -I_\tau = \pm
\frac{1}{2}\sin\theta_{13}$. Since these invariants are independent of
$\theta_X$, one can use the leading order values of $\theta_{12}$ to obtain
information on $\delta_{\rm CP}$. Theses are determined by the choice of $T_l$
and $T_\nu$. If $c_{12}s_{12}\neq 0$ at the leading order, then the above
equations predict $\beta_1=0$ and $\delta_{\rm CP}=\pm \frac{\pi}{2}$. On the
other hand if $c_{12}s_{12} = 0$ at the leading order than one gets
$\sin(\delta_{\rm CP} \pm \frac{\beta_1}{2})=\pm 1$. It is thus expected that
small perturbations will stabilize $\delta_{\rm CP}$ around the values obtained
in these two cases depending on the choice of the residual symmetries. Examples
of specific perturbations doing this have been considered in
\cite{Hernandez:2013vya}. Also general perturbations to the $U_{\rm PMNS}$
matrix obtained in case of the $A_5$ group were numerically analyzed in \cite{Joshipura:2014qaa} and
$\delta_{\rm CP}$ was found to be near $\pm \frac{\pi}{2}$ for the choices of
$T_l$ and $T_\nu$ made there. We shall give here an explicit model where one gets
the same result after perturbations. 
\item The third column of $U$ is not
affected by arbitrariness in the choice of $\theta_{12}$ and the values of
$\theta_{13}$ is uniquely fixed by the choice of $T_\nu$ and $T_l$. We consider
leading order prediction of $\theta_{13}$ for DSG of $O(3)$ in the next section
concentrating mainly on $A_5$ . \end{itemize}

\section{Examples of generalized \mt symmetry and $A_5$} 
\label{example} 
The groups $S_3,~D_N,~A_4,~S_4$ and $A_5$ are the only finite DSG of $O(3)$. Of
these only $A_4,~S_4$ and $A_5$ posses faithful three dimensional irreducible
representations. Any choice of residual symmetries within them consistent with
the previous discussion would lead to prediction Eq. (\ref{gen-pred}). The
mixing angle predictions for $A_5$ group have already been studied
\cite{Everett:2008et,Feruglio:2011qq,Cooper:2012bd,Ding:2011cm} in case of the
non-degenerate neutrinos. One gets either vanishing or large $\theta_{13}$ at the leading order in
this case. The same holds for the groups $A_4$ and $S_4$
even in case of the partially degenerate spectrum. The group $A_5$ provides only
non-trivial example which gives a non-zero $\theta_{13}$ close to its
experimental value if two of the neutrinos are degenerate. We discuss this case
explicitly and enumerate all the residual symmetries within $A_5$ giving
generalized \mt symmetry.

The $A_5$ group has sixty elements which are generated using $E$, $F$ and $H$
where 
\be \label{def} H=1/2\left(\ba{ccc} -1&\mu_ -&\mu_+\\ \mu_-&\mu_+&-1\\
\mu_+&-1&\mu_-\\ \ea \right) ~;~~~E=\left(\ba{ccc} 0&1&0\\ 0&0&1\\ 1&0&0\\ \ea
\right) ~;~~~F=\left(\ba{ccc} 1&0&0\\ 0&-1&0\\ 0&0&-1\\ \ea \right)~,\ee 
with
$\mu_\pm=1/2(-1\pm \sqrt{5})$. We list all the sixty elements in terms of $E$,
$F$, $H$ defined above in the Appendix. Properties of $A_5$ group has been
studied earlier in \cite{Everett:2008et,Feruglio:2011qq,Cooper:2012bd} and
reference \cite{Ding:2011cm} also gives list of all elements using different
matrices. We have defined them in a way which makes the appearance of the
generalized \mt symmetry for $A_5$ explicit.

We divide the sixty elements into four categories: (i) An identity element, (ii)
the 15 elements of order 2 to be collectively called $O_2$. The character $\chi$ of
these elements is $-1$, (iii) the 20 elements of order 3 to be called $O_3$, all
with $\chi=0$ and (iv) 24 elements of order 5 collectively called $O_5$. The 12
of these have $\chi=-\mu_+ $ and another 12 have $\chi=-\mu_-$. All these
elements and their diagonalizing matrices are listed in Table \ref{elements} in
Appendix. Following Eq. (\ref{ev-su3}), we find that all the elements in
category $O_3$ and $O_5$ have one real and two complex conjugate eigenvalues.
Thus there are 44 elements belonging to $O_3$ and $O_5$ which qualify to be the
residual symmetry $T_\nu$, $T_l$ of neutrinos and the charged leptons
respectively. The 15 elements in $O_2$ contain five distinct $Z_2\times Z_2$
subgroups which can be used as residual symmetry of $M_\nu$ in case of the
non-degenerate spectrum. For each of these five choices, there exists 44 $T_l$
giving generalized \mt or $e$-$\tau$ or $e$-$\mu$ symmetry. The last two can be
converted to $U_{\rm PMNS}$ satisfying Eq. (\ref{gen-pred}) after proper
reordering in the columns of $T_l$. If any of the five $Z_2\times Z_2$ group is
used as residual symmetry of $M_\nu$ and any of 24 elements in class $O_5$ as
$T_l$ then one gets the following $|U_{\rm PMNS}|$:
\be \label{nondeg}
|U_{\rm
PMNS}|= \left(
\begin{array}{ccc}
 0.8507 & 0.5257 & 0 \\
 0.3717 & 0.6015 & 0.7071 \\
 0.3717 & 0.6015 & 0.7071 \\
\end{array}
\right)~\ee
or
matrix which
differs from above by reordering of row and columns. This matrix has the
property of golden ratio prediction \cite{Kajiyama:2007gx} for the solar mixing
angle $\sin^2\theta_{12}=0.276$. It however predicts $\sin^2 \theta_{13}=0$. This case provides a
good zeroeth order approximation and it has already been discussed in
\cite{Kajiyama:2007gx,Everett:2008et,Feruglio:2011qq,Ding:2011cm,Cooper:2012bd}.
If one chooses any of 20 elements in $O(3)$ as $T_l$ then one gets generalized
\mt symmetry but the resulting form of $|U_{\rm PMNS}|$ differs significantly from the
observed one.

In case of the partially degenerate neutrino spectrum, one has the choice of 44 elements as residual
symmetries of $M_\nu$ and $M_l$ consistent with generalized \mt. The structure
of the PMNS matrix follows from the basic structure of $U_l$, $U_\nu$. In
particular, one gets from Eq. (\ref{gen-vl}) and Eq. (\ref{vnu}) 
$$
\sin^2 \theta_{13}=\left|\sum_{i} x_i u_i\right|^2~, $$ 
where $x_i$ ($u_i)$ denotes the
eigenvector of
$T_l$ ($T_\nu$) corresponding to the eigenvalue 1. This can be determined from
the structure of the elements $O_2$ and $O_5$ as given in the Appendix. All
possible values of $\theta_{13}$ obtained in this way are give by
$$\sin^2\theta_{13} = \{0.035,~0.111,~0.2,~0.556,~0.632\}~.$$ Similar exercise
in case of the $A_4$ and $S_4$ groups gives: \beqa \label{}
A_4~~&:&~~\sin^2\theta_{13} = 0.111~; \nonumber \\ S_4~~&:&~~\sin^2\theta_{13} =
\{ 0,~0.111,~0.333\}~.\eeqa The same results also follow from
\cite{Joshipura:2014qaa} in which an extensive analysis was performed on several
discrete subgroups of $SU(3)$ which can lead to the appropriate symmetries for
degenerate solar pair. The numerical results presented in Table I in
\cite{Joshipura:2014qaa} shows that among all the analyzed groups, the only group with prediction maximal
$\theta_{23}$ and $\delta_{\rm CP}$ for $0<\sin^2\theta_{13}<0.05$ is $A_5$ or
the group which contains it as a subgroup, for example $\Sigma(1080)$.

Of all the predicted values, $\sin^2\theta_{13}=0.035$ can be considered close
to experiments which can be brought within 3$\sigma$ limit of the experimental
value with relatively small corrections. This value is obtained if $T_l$ belongs
to $O_5$ and $T_\nu$ to $O(3)$ or vice versa. There exists more than one
structures of $|U_{\rm PMNS}|$ corresponding to the same value of
$s_{13}^2$. We
note here two qualitatively different cases.

If $T_l=T$ and $T_\nu= E^{-1}AE$ then one gets 
$$|U_{\rm PMNS}|=\left(
\begin{array}{ccc} 0.8507 & 0.4911 & 0.1876 \\ 0.3717 & 0.616 & 0.6946 \\ 0.3717
& 0.616 & 0.6946 \\ \end{array} \right)$$ 
upto a rotation by an angle $\theta_{X}$ in the 12 plane, where $A$, $T$ are
defined in the Appendix. The
same $T_l$ but $T_\nu=AEA^{-1}$ gives instead $$|U_{\rm PMNS}|=\left(
\begin{array}{ccc} 0.9822 & 0 & 0.1876 \\ 0.1326 & 0.7071 & 0.6946 \\ 0.1326 &
0.7071 & 0.6946 \\ \end{array} \right)$$ In the first case, $c_{12}s_{12}$ is
non-zero at the leading order. Then invariants given in Eq. (\ref{invariant})
lead to $\beta_1=0$, $\delta_{\rm CP}=\pm \frac{\pi}{2}$. In the second case,
$c_{12}s_{12}=0$ and one gets $\sin(\delta_{\rm
CP}\pm \beta_1/2)=\pm 1$. The small perturbations are then required to fix
$\theta_{12}$ to its experimental value and to generate splittings in the solar
pair. Such perturbations would also fix $\delta_{\rm CP}$ close to the
values around this.
\section{An $A_5$ model} \label{model} We now provide explicit model in
which
the results of previous section can be realized. The model is very similar to
the one presented in \cite{Varzielas:2013hga}. Major difference being a
different vacuum alignment and the form of the charged lepton mass matrix. The
group $A_5$ has ${\bf 1}$, ${\bf 3_1}$, ${\bf 3_2}$, ${\bf 4}$ and
${\bf 5}$ dimensional irreps where ${\bf 3_1}$ and ${\bf 3_2}$ are
non-equivalent irreps. The model is supersymmetric with the 
three generations of leptons $l_L$ and $l^c$ both transforming  as ${\bf 3_1}$ under
$A_5$
as in \cite{Varzielas:2013hga}. It follows from the product $${\bf 3_1}\times
{\bf 3_1}=({\bf 1}+{\bf 5})_{\rm sym.}+{\bf 3_1}_{\rm antisym.}$$ that symmetric
neutrino masses can arise from ${\bf 1}+{\bf 5}$ and the charged lepton masses
can arise from all three irreps. Accordingly, we introduce two flavons, a 5-plet
$\phi_\nu$ and a singlet $s_\nu$ to generate neutrino masses. The Higgs doublets
of the minimal supersymmetric standard model, $H_u$ and $H_d$, are singlet of
$A_5$. We introduce a weak triplet $\Delta$ as an $A_5$ singlet. The relevant
superpotential is: \be \label{wnu} W_\nu= \frac{1}{2 \Lambda} l_L^T \Delta
l_L(h_{s\nu } s_\nu +h_{5\nu} \phi_\nu)~.\ee The charged lepton masses are
generated by three additional flavons, a singlet $s_l$, a 5-plet $\phi_l$ and a
triplet $\chi_l$. The corresponding superpotential is \be \label{wl} W_l=
\frac{1}{ \Lambda} l_L H_d l^c (h_{sl} s_l +h_{5l} \phi_l+h_{3l} \chi_l)~.\ee

Among the various possible choices of the residual symmetries given in the
Appendix, we specialize to a particular choice with $T_l=E$ and $T_\nu=f_2 T
f_2$. A hermitian combination of the charged lepton mass matrix $M_lM_l^\dagger$
invariant under $T_l$ results if the vacuum expectation values (VEV)
$\vev{\chi_l}$ and $\vev{\phi_{l}}$ satisfy 
\be \label{lvac}
T_l(3)\vev{\chi_l}=\vev{\chi_l}~,~~~ T_l(5)\vev{\phi_l}=\vev{\phi_l}~,\ee where
$T_l(3)$ $(T_l(5))$ denotes the matrices corresponding to the ${\bf 3_1}$ $({\bf
5})$ representation. The $T_l(3)=E$ and $T_l(5)$ is given 
\cite{Varzielas:2013hga} by:
\be T_l(5)=\left( \begin{array}{ccccc} 0 & 1 &
0 & 0 &
0 \\ 0 & 0 & 1 & 0 & 0 \\ 1 & 0 & 0 & 0 & 0 \\ 0 & 0 & 0 & -\frac{1}{2} &
-\frac{\sqrt{3}}{2} \\ 0 & 0 & 0 & \frac{\sqrt{3}}{2} & -\frac{1}{2} \\
\end{array} \right)~. \ee Denoting $\vev{\chi_l}=(\chi_1,\chi_2,\chi_3)^T$ and
$\vev{\phi_l}=(q_1,q_2,q_3,q_4,q_5)^T$, Eqs. (\ref{lvac}) are solved by \be
\label{soll} v_1=v_2=v_3\equiv v_l~,~~ q_1=q_2=q_3\equiv q_l~~{\rm
and}~~q_4=q_5=0~.\ee 
Inserting this solution in the superpotential in Eq.
(\ref{wl}) leads to a charged lepton mass matrix 
\be \label{ml} M_l=\left(
\ba{ccc} m_0&m_1-m_2&m_1+m_2\\ m_1+m_2&m_0&m_1-m_2\\ m_1-m_2&m_1+m_2&m_0\\ \ea
\right)~. \ee 
The $M_lM_l^\dagger$ is diagonalized by the matrix $U_\omega$ which
also diagonalizes the corresponding symmetry generator $T_l(3)=E$. Explicitly, 
\be U_\omega^{\dagger} M_lM_l^\dagger
U_\omega={\rm Diag.} (m_e^2,m_\mu^2,m_\tau^2)~.\ee with \beqa m_e^2&=& |m_0 + 2
m_1|^2~,\nonumber \\ m_\mu^2&=&|m_0-m_1- \sqrt{3} i ~m_2|^2~,\nonumber \\
m_\tau^2&=&|m_0-m_1 + \sqrt{3} i ~m_2|^2~\eeqa
Here $m_0$ can be taken real
without loss of generality. Note that the electron mass given above corresponds
to the eigenvector $(1,1,1)^T$ of $U_\omega$. This has to be identified as the
first column of $U_l$ in order to get the \mt symmetry as already mentioned. The
remaining two eigenvalues can be identified with muon and tau lepton masses and
can be interchanged. The contributions labeled by $m_0$, $m_2$, $m_1$ arise from
the VEVs of singlet, triplet and the 5-plet. The $M_l$ is symmetric in the
absence of triplet. In this case, $T_l$ invariance implies two degenerate
charged leptons. Thus a large triplet contribution $m_2$ is essential to split
the muon and tau lepton masses. Moreover, simultaneous presence of $m_0$ and $m_1$
is also required to suppress the electron mass. But given all the three
contributions, one can fit the charged lepton masses with appropriate choice of
parameters.

Neutrino masses follow analogously from Eq. (\ref{wnu}). In order to get
degeneracy, we impose the residual symmetry $T_{\nu}=f_2 T f_2$ and require that
\be \label{nuvac} T_\nu(5)\vev{\phi_\nu}=\vev{\phi_\nu}~,\ee where $T_{\nu}(5)$
can be shown to be\footnote{This is determined by noting that the presentations
$a,b,c$ introduced in \cite{Varzielas:2013hga} are given in terms of our
presentation as $f_3=a$, $E=b$, $ f_1= b^2 a b$ and $ H=b c$.} 
\be
T_\nu(5)=\left( \begin{array}{ccccc} -\frac{1}{2} & 0 & \frac{1}{2} & \frac{1}{2
\sqrt{2}} &\frac{\sqrt{3}}{2\sqrt{2}} \\ 0 & \frac{1}{2} & \frac{1}{2} &
-\frac{1}{\sqrt{2}} & 0 \\ \frac{1}{2} & -\frac{1}{2} & 0 & -\frac{1}{2
\sqrt{2}} & \frac{\sqrt{3}}{2\sqrt{2}} \\ -\frac{1}{2 \sqrt{2}} &
-\frac{1}{\sqrt{2}} & \frac{1}{2 \sqrt{2}} & -\frac{1}{4} & -\frac{\sqrt{3}}{4}
\\ -\frac{\sqrt{3}}{2\sqrt{2}} & 0 & -\frac{\sqrt{3}}{2\sqrt{2}} &
-\frac{\sqrt{3}}{4} & \frac{1}{4} \\ \end{array} \right)~.\ee 
Let
$\vev{\phi_\nu}=(p_1,p_2,p_3,p_4,p_5)^T$. A solution for Eq. (\ref{nuvac}) is
given by 
\be \label{ps} p_1=p_3=0~,~p_2=-\sqrt{2}
p_4~,~p_5=-\frac{p_4}{\sqrt{3}}~.\ee Inserting these in the neutrino
superpotential leads to a neutrino mass matrix 
\be\label{mnu0} M_{0\nu}=\left(
\ba{ccc} m_{0\nu }-\frac{m_{1\nu}}{3}(\mu_+-\mu_-)&0&0\\ 0&m_{0 \nu
}-\frac{m_{1\nu}}{3}(\mu_- -1)&-m_{1\nu}\\ 0&-m_{1\nu}&m_{0 \nu
}-\frac{m_{1\nu}}{3}(1-\mu_+)\\ \ea \right)~.\ee
As a consequence of the
residual symmetry, one gets two degenerate neutrinos with a mass
$m_{0\nu}-\frac{m_{1\nu}}{3}(\mu_+-\mu_-)$ and the third mass is given by
$m_{0\nu}+\frac{2 m_{1\nu}}{3}(\mu_+-\mu_-)$. The lower $2\times 2$ block of
$M_\nu$ is diagonalized by a rotation matrix with an angle $\theta$ given by:
$$\tan\theta=-\mu_-.$$ The full PMNS matrix at the leading order is thus given
by 
\be \label{upmns0} U_{0}\equiv U_\omega^{\dagger} R_{23}(\theta)=
\frac{1}{\sqrt{3}}\left( \ba{ccc} 1&c_\theta+s_\theta&c_\theta-s_\theta \\
1&c_\theta\omega^2+s_\theta\omega&c_\theta \omega-s_\theta\omega^2 \\
1&c_\theta\omega+s_\theta\omega^2&c_\theta \omega^2-s_\theta\omega \\ \ea
\right) ~,\ee 
where $c_\theta=\cos\theta,s_\theta=\sin\theta$. The generalized
\mt symmetry is apparent from the
above. Moreover, \be\label{s13model0}
s_{13}^2=\frac{1}{3}(c_\theta-s_\theta)^2=\frac{1}{3} \left(1+
\frac{2\mu_-}{1+\mu_-^2}\right)\approx 0.035~\ee as would be expected from the
specific choice of the residual symmetry made in this example.

The above zeroth order prediction would get modified from the perturbations
which are required to split the degenerate states, fix the solar angle and to
change the zeroth order predictions for the mixing angles $\theta_{13}$ and
$\theta_{23}$. Effects of general perturbations were studied in \cite{Joshipura:2014qaa}
in the context of
$A_5$ symmetry with a slightly different choice of the residual symmetry which
also leads to the same zeroth order predictions as here. It was found that small
perturbations can cause significant changes in $\theta_{13}$ as required
experimentally and relatively small perturbations in the zeroth order values of
$\theta_{23}$ and the maximal CP violating phase. Moreover, all three neutrinos
are required to be quasidegenerate in order to reproduce all the mixing angles
correctly as long as perturbations are smaller than $\le 5\%$. The analysis in
\cite{Joshipura:2014qaa} was for the most general possible perturbations. In the
context of specific models, such perturbations can arise from the non-leading
higher order terms in the Yukawa superpotential which directly correct the
leptonic mass matrices and/or from the Higgs potential which may perturb the
Higgs vacuum expectation values from the symmetric choice. Let us consider
effect of a simple but interesting perturbation in the latter category. Assume
that the perturbations change one of the VEVs given in Eq. (\ref{ps}),
namely  $p_2\rightarrow p_2(1+\epsilon)$. Similar perturbations in the
VEV of other component would also arise in general but as we discuss here, this
perturbation alone has interesting consequences. The zeroth order mass matrix in
Eq. (\ref{mnu0}) now gets changed to 
\be\label{mnucorr} M_\nu=\left( \ba{ccc}
m_{\nu 0}-\frac{m_{1\nu}}{3}(\mu_+-\mu_-)&0&0\\ 0&m_{\nu
0}-\frac{m_{1\nu}}{3}(\mu_- -1)&-m_{1\nu}(1+\epsilon)\\
0&-m_{1\nu}(1+\epsilon)&m_{\nu 0}-\frac{m_{1\nu}}{3}(1-\mu_+)\\ \ea \right)~.\ee
The above perturbed matrix is also diagonalized by a rotation in the 2-3
plane
but with a slightly different $\theta$ which is now given by 
$$\tan\theta\approx
-\mu_- \left(1-\frac{\epsilon}{\sqrt{5}} \right)+{\cal O} (\epsilon^2)~.$$ This
changes the zeroth order prediction of the mixing angle $\theta_{13}$ and Eq.
(\ref{s13model0}) gets replaced by
\be \label{news13}
s_{13}^2=\frac{1}{3}(c_\theta-s_\theta)^2=
\frac{\mu_+^2}{3(1+\mu_-^2)}-\frac{2\epsilon}{3\sqrt{5}}\frac{\mu_-^2}{
(1+\mu_-^2)^2}+{\cal O} (\epsilon^2)~.\ee
Thus the appropriate choice of
perturbation can be used to get agreement with experiments. The other major
effect of $\epsilon$ is to split the degenerate pair and induce the solar scale:
$$\frac{\Dms}{\Dma} \equiv \frac{\Delta m^2_{21}}{\Delta
m^2_{31}}=\frac{4\epsilon}{5}\left(\frac{3\sqrt{5}m_{0\nu}-5 m_{1\nu}}{6\sqrt{5}
m_{0\nu}+5 m_{1\nu}}\right)+{\cal O} (\epsilon^2) $$

The overall effect of the perturbation is best appreciated by going to the
flavour basis with $M_lM_l^\dagger$ diagonal. In this basis
\beqa \label{mnuf}
M_{\nu f}&\equiv& U_\omega^{T} M_\nu U_\omega~\nonumber \\ &=&\left( \ba{ccc}
m_{0\nu}-\frac{2}{3} m_{1\nu} (1+\epsilon)&-\frac{m_{1\nu}}{3}(\mu_+
+\omega^2-\epsilon)&-\frac{m_{1\nu}}{3}(\mu_+ +\omega-\epsilon)\\
-\frac{m_{1\nu}}{3}(\mu_+ +\omega^2-\epsilon)&-\frac{m_{1\nu}}{3}(1-\mu_-
-\omega^2+2\epsilon)&m_{0\nu}+\frac{m_{1\nu}}{3}(1+\epsilon)\\
-\frac{m_{1\nu}}{3}(\mu_+
+\omega-\epsilon)&m_{0\nu}+\frac{m_{1\nu}}{3}(1+\epsilon)&-\frac{m_{1\nu}}{3}(1-\mu_-
-\omega+2\epsilon)\\ \ea\right) \eeqa
The interesting features of this matrix
are: \begin{itemize} \item Elements of $M_{\nu f}$ satisfy $$ \sum_{i}(M_{\nu
f})_{e i}=\sum_{i}(M_{\nu f})_{\mu i}=\sum_{i}(M_{\nu f})_{\tau i}~.$$ This
condition implies that one of the column vectors of $U_{\rm PMNS}$ has a
tri-maximal form as is the case with the zeroth order mixing matrix, Eq.
(\ref{upmns0}). Thus one gets the prediction
$\sin^2\theta_{12}\cos^2\theta_{13}=\frac{1}{3}$ if perturbation makes the state
with an eigenvector corresponding to the first column in Eq. (\ref{upmns0})
heavier compared to the second degenerate state. Perturbation in this case does
not change the zeroth order solar angle but it stabilizes it to that value by
splitting the degenerate states. \item If parameters $m_{0\nu}$, $m_{1\nu}$ and
$\epsilon$ are real then the $M_{\nu f}$ satisfies $(M_{\nu f})_{12}=(M_{\nu
f})_{13}^*$ and $(M_{\nu f})_{22}=(M_{\nu f})_{33}^*$. Thus $M_{\nu f}$
simultaneously enjoys the $Z_2\times Z_2$ symmetries corresponding to a
tri-maximal solar angle and generalized \mt as envisaged and studied in
\cite{Gupta:2011ct}. In particular, one gets the maximal atmospheric angle and
the maximal CP violating phase as predictions even after perturbation.
\end{itemize}

As an example, we give a set of specific values of $\epsilon$, $m_{0\nu}$,
$m_{1\nu}$ determined numerically which fit the experimental values: $$
m_{0\nu}=0.025~{\rm eV},~~m_{1\nu}=0.016~{\rm eV},~~\epsilon= 0.228$$ This
gives the following
mixing matrix 
$$|U_{\rm PMNS}|^2= \left( \begin{array}{ccc} 0.6421 & 0.3333 &
0.0246 \\ 0.179 & 0.3333 & 0.4877 \\ 0.179 & 0.3333 & 0.4877 \\ \end{array}
\right)~.$$ 
corresponding to
$$\sin^2\theta_{12}\cos^2\theta_{13}=\frac{1}{3}~,~\sin^2\theta_{23}=\frac{1}{2}~,~\sin^2\theta_{13}=0.0246~.$$
The $\delta_{\rm CP}$ gets stabilized to the value $-\pi/2$. The neutrino masses
giving correct $\Dms$ and $\Dma$ are determined by the above values of
parameters as 
$$ m_{\nu_1}=0.0097~{\rm eV},~~m_{\nu_2}=0.0131~{\rm
eV},~m_{\nu_3}=0.0522~{\rm eV}.$$ 
The maximality of $\theta_{23}$ can be changed
by introducing small imaginary parts in parameters but the tri-maximal value of
$\theta_{12}$ remains unchanged. Small deviations can be introduced by
perturbing other component of the VEVs or by perturbing the charge lepton mass
matrix. Since general perturbations are already studied in
\cite{Joshipura:2014qaa}, we shall not pursue them further.

\section{Summary} The generalized \mt symmetry of the leptonic mixing matrix is
known to predict maximal atmospheric mixing angle and maximal Dirac CP violation
in case of nonzero $\theta_{13}$. Both these predictions are consistent with the
current experimental observations within $3\sigma$ and their future precision
measurements will reveal weather such a symmetry is indeed realized in nature in
its exact form. It is therefore interesting to explore the symmetries of the
leptons which lead to generalized \mt symmetry in the lepton mixing predicting
such special values of $\theta_{23}$ and $\delta_{\rm CP}$.

Assuming the Majorana neutrinos, we have shown in this paper that generalized \mt
symmetry naturally follows if the symmetry group $G_f$  of  leptons,
is a discrete subgroup of $O(3)$. It is required that the $G_f$ is broken
into
$Z_m$ with $m \ge 3$ as the residual symmetry of the charged lepton mass
matrix.
The residual symmetry of the Majorana neutrino mass matrix can be either (a)
$Z_2 \times Z_2 \in G_f$ or (b) $Z_n \in G_f$ with $n \ge 3$. The possibility
(a) leads to three non-degenerate neutrinos while one obtains two of the three
neutrinos degenerate in the case (b). The possible candidates of $G_f$ are only
$A_4$, $S_4$ and $A_5$ which can predict all the three mixing angles at the
leading order. Among these, only $A_5$ predicts $\theta_{13}$ very close to its
experimentally observed value in the case of two degenerate neutrinos which are
identified with the solar pair. The corrections to the leading order neutrino
mass matrix are needed to generate viable $\theta_{13}$, $\theta_{12}$ and the
solar mass difference. We have discussed in detail the group $A_5$ in the context of
generalized \mt symmetry and provided an explicit model in which the leading
order predictions are realized. We have also discussed the perturbations which lead
to the realistic neutrino masses and mixing angles while maintaining the
predictions $\theta_{23}=\pi/4$ and $\delta_{\rm CP} = \pm \pi/2$.

Some example ansatz and symmetries of neutrino mass matrix leading to the
generalized \mt symmetry have already been discussed in the literature. Our
findings of an emergence of generalized \mt symmetry from the discrete subgroups
of $O(3)$ are more general and they accommodate some of the symmetries and
models
proposed in literature to obtain $\theta_{23}=\pi/4$ and $\delta_{\rm CP} = \pm
\pi/2$. In particular, we have shown that the generalized \mt symmetry in the lepton
mixing can follow without imposing \mt symmetry and/or CP on the neutrino mass
matrix. The \mt symmetry with CP conjugation is realized in our approach only
accidentally when an additional assumption is made on the free parameters.

\begin{acknowledgments} A.S.J. thanks the Department of Science and Technology,
Government of India for support under the J. C. Bose National Fellowship
programme, grant no. SR/S2/JCB-31/2010. K.M.P. thanks the Department of Physics
and Astronomy of the University of Padova for its support. He also acknowledges
partial support from the European Union FP7 ITN INVISIBLES (Marie Curie
Actions, PITN-GA-2011-289442). \end{acknowledgments}

\section{Appendix} We list all the sixty elements belonging to $A_5$ in terms of
their presentation matrices $E$, $F$ and $H$ defined in Eq. (\ref{def}). For
brevity, we have defined the following matrices which are used to
characterize
various elements. 
$$
f_1=F~,~~f_2=Ef_1E^{-1}~,~~f_3=Ef_2E^{-1},~~T=f_1EH~,~~A=Hf_1~.$$ 
The elements are listed in Table \ref{elements}. 
\begin{table}[!ht]
\label{generators} \begin{center} 
\begin{math} 
\begin{tabular}{ccc} \hline
\hline (Set, Order) & Set of elements & Diagonalizing matrix\\ \hline
\multirow{3}{*}{($O_2,~2)$}
&$f_a$ & $I$ \\
& $H$ & $U_H$ \\
& $ f_a Hf_a$ & $f_aU_H $\\
& $EHE^{-1}$ & $EU_H$\\
& $E^{-1}HE$ & $E^{-1}U_H$\\ 
& $Ef_a H f_aE^{-1}$ & $Ef_aU_H$ \\
& $E^{-1}f_aHf_aE$ & $E^{-1}f_aU_H$\\ 
\hline
\multirow{7}{*}{($O_3,~3$)}&$E, ~E^2=E^{-1}$ & $U_\omega$\\ & $f_a E
f_a,~f_aE^{-1}f_a$ & $f_a U_\omega$\\ & $A,~A^2=A^{-1}$ & $U_A$\\ & $EAE^{-1},
~EA^2E^{-1}$ & $EU_A$\\ & $E^{-1}AE, ~E^{-1}A^2E$ & $E^{-1}U_A$\\ &
$AEA^{-1},~AE^{-1}A^{-1}$ & $AU_\omega$\\ & $Af_{2,3}Ef_{2,3}A^{-1},
~Af_{2,3}E^{-1}f_{2,3}A^{-1}$ & $A f_{2,3} U_\omega$\\ 
\hline
\multirow{7}{*}{$(O_5,~5)$} & $
T^p$ & $U_T$ \\ 
& $f_2T^pf_2$ & $f_2 U_T$ \\
& $ET^pE^{-1}$ & $E U_T$ \\
& $E^{-1}T^p E$ & $E^{-1}U_T$ \\
& $Ef_2T^pf_2E^{-1}$ & $Ef_2U_T$\\ &
$E^{-1}f_2T^pf_2E$ & $E^{-1}f_2U_T$\\
\hline
\hline 
\end{tabular}
\end{math} \end{center}
\caption{List of all the non-trivial elements of $A_5$. The last column gives
the list of diagonalizing matrices for the corresponding elements which are used as the residual symmetries of neutrino and/or charged
lepton mass matrices. The $T^p$ collectively denotes a list of four elements
$T^p=(T,T^2,T^3,T^4)$ while $a=1,2,3$~.} \label{elements} \end{table}
Here $U_\omega$ diagonalizes $E,E^2$ and  is  defined in Eq. (\ref{uw}). The unitary matrices
$U_A$, $U_T$ and $U_H$ respectively diagonalize $(A,A^2)$, $T^p$ and $H$ and are given by
\beqa\label{uAuT} U_A&=&\left( \ba{ccc}
\frac{i}{\sqrt{2}}&-\frac{i}{\sqrt{2}}&0\\
\frac{\mu_-}{\sqrt{6}}&\frac{\mu_-}{\sqrt{6}}&-\frac{\mu_+}{\sqrt{3}}\\
\frac{\mu_+}{\sqrt{6}}&\frac{\mu_+}{\sqrt{6}}&\frac{\mu_-}{\sqrt{3}}\\
\ea\right)~,~~U_T=\frac{1}{\sqrt{2}}\left( \ba{ccc} 1&1&0\\ x \mu_ -&x^*
\mu_-&-\frac{\sqrt{2}\mu_-}{(1+\mu_-^2)^{1/2}}\\ -x (\mu_- -1)&-x^* (\mu_-
-1)&-\frac{\sqrt{2}}{(1+\mu_-^2)^{1/2}}\\ \ea\right)~\nonumber \\
U_H &=&\left(
\ba{ccc}
0&-\frac{\sqrt{3}}{2}&\frac{1}{2}\\
\frac{\mu_+}{\sqrt{3}}&\frac{\mu_-}{2\sqrt{3}}&\frac{\mu_-}{2}\\
-\frac{\mu_-}{\sqrt{3}}&\frac{\mu_+}{2\sqrt{3}}&\frac{\mu_+}{2}\\
\ea \right) U_{12}~. \eeqa
Here $U_{12}$ denotes an arbitrary unitary rotation in the $12$ plane arising due to
degeneracy in two of the eigenvalues of $H$ and
$x=\frac{\lambda+1}{\lambda-1}$ with
$\lambda=\frac{1}{2}(\mu_-+i\sqrt{4-\mu_-^2})$. All the non-trivial elements of
$A_5$ given in the Table are expressed in the form $QPQ^{-1}$ with
$P=E,E^2,A,T^p,H$ and $Q=I,E,E^2,f_a,Ef_a,E^2f_a,Af_a$. This simplifies
diagonalization of
all the elements since $U_{QPQ^{-1}}=QU_P$. This allows one in principal to
calculate all possible $U_{\rm PMNS}$ in $A_5$ analytically in terms the
diagonalizing matrices of $E,A,T,H$. In particular, matrices diagonalizing $A$,
$T$, $E$ and therefore all elements in $O_5$, $O_3$ are seen to have
\mt symmetric form given in Eq. (\ref{gen-vl}).
\bibliography{ref-a5.bib} \end{document}